\documentclass[journal]{IEEEtran}
\usepackage[utf8]{inputenc}
\usepackage{cite}
\usepackage{graphicx}
\usepackage{amsmath}
\usepackage{mathrsfs,amsthm,amsmath,amssymb}
\usepackage{textcomp}
\usepackage{graphicx,multirow}
\usepackage{balance}
\usepackage{cite}
\usepackage{float}
\usepackage{gensymb}
\usepackage{eso-pic}
\usepackage{algorithm}
\usepackage{algorithmic}
\usepackage[hidelinks]{hyperref}
\usepackage{academicons}
\usepackage{xcolor}
\makeatletter
\newcommand\subparagraph{%
  \@startsection{subparagraph}{5}
  {\parindent}
  {3.25ex \@plus 1ex \@minus .2ex}
  {-1em}
  {\normalfont\normalsize\bfseries}}
\makeatother
\usepackage{titlesec}
\let\subparagraph\relax

\usepackage{titlesec}
\usepackage{scalerel}
\usepackage{tikz}
\usepackage{subcaption}
\usepackage[english]{babel}

\newcommand{\bs}{\boldsymbol}
\newcommand{\mb}{\mathbf}

\DeclareMathOperator{\snr}{SNR}

\DeclareMathOperator{\db}{dB}

\DeclareMathOperator{\de}{d}
\DeclareMathOperator{\ew}{elsewhere}

\usetikzlibrary{svg.path}

\definecolor{orcidlogocol}{HTML}{A6CE39}
\tikzset{
  orcidlogo/.pic={
    \fill[orcidlogocol] svg{M256,128c0,70.7-57.3,128-128,128C57.3,256,0,198.7,0,128C0,57.3,57.3,0,128,0C198.7,0,256,57.3,256,128z};
    \fill[white] svg{M86.3,186.2H70.9V79.1h15.4v48.4V186.2z}
                 svg{M108.9,79.1h41.6c39.6,0,57,28.3,57,53.6c0,27.5-21.5,53.6-56.8,53.6h-41.8V79.1z M124.3,172.4h24.5c34.9,0,42.9-26.5,42.9-39.7c0-21.5-13.7-39.7-43.7-39.7h-23.7V172.4z}
                 svg{M88.7,56.8c0,5.5-4.5,10.1-10.1,10.1c-5.6,0-10.1-4.6-10.1-10.1c0-5.6,4.5-10.1,10.1-10.1C84.2,46.7,88.7,51.3,88.7,56.8z};
  }
}

\newcommand\orcidicon[1]{\href{https://orcid.org/#1}{\mbox{\scalerel*{
\begin{tikzpicture}[yscale=-1,transform shape]
\pic{orcidlogo};
\end{tikzpicture}
}{|}}}}

\newcommand{\revisioncolor}{black}
\newcommand{\omissioncolor}{black}

\graphicspath{ {Images/} }
\IEEEoverridecommandlockouts

\begin{document}
\setlength{\parskip}{5pt}
\setlength{\abovedisplayskip}{5pt}
\setlength{\belowdisplayskip}{5pt}
\title{AoA-Based Pilot Assignment in Massive MIMO Systems Using Deep Reinforcement Learning }

\author{Yasaman~Omid$^\text{\orcidicon{0000-0002-5739-8617}}$, Seyed MohammadReza~Hosseini$^\text{\orcidicon{0000-0002-4189-3533}}$, Seyyed MohammadMahdi~Shahabi$^\text{\orcidicon{0000-0003-2873-4156}}$,\\
Mohammad~Shikh-Bahaei$^\text{\orcidicon{0000-0001-7450-7574}}$,
Arumugam~Nallanathan$^\text{\orcidicon{0000-0001-8337-5884}}$
 
 \thanks{Yasaman Omid and Arumugam Nallanathan  are with the School of Electronic Engineering and Computer Science, Queen Mary University of London, U.K. (e-mail: y.omid@qmul.ac.uk; a.nallanathan@qmul.ac.uk).} 
 
 \thanks{Seyyed MohammadReza Hosseini is with the Department of Electrical Engineering, K. N. Toosi University of Technology, Tehran, Iran (e-mail: s.m.hosseini@ee.kntu.ac.ir).}

\thanks{Seyyed MohammadMahdi Shahabi and Mohammad Shikh-Bahaei are with the Department of Engineering, Kings College London, U.K. (e-mail: mahdi.shahabi@kcl.ac.uk;  m.sbahaei@kcl.ac.uk).}
 }

\maketitle
\begin{abstract}
In this paper, the problem of pilot contamination in a multi-cell massive multiple input multiple output (M-MIMO) system is addressed using deep reinforcement learning (DRL). To this end, a pilot assignment strategy is designed that adapts to the channel variations while maintaining a tolerable pilot  contamination effect. Using the angle of arrival (AoA) information of the users, a cost function, portraying the reward, is presented, defining the pilot contamination effects in the system.
Numerical results illustrate that the DRL-based scheme is able to track the changes in the environment, learn the near-optimal pilot assignment, and achieve a close performance to that of the optimum pilot assignment performed by exhaustive search, while maintaining a low computational complexity.   
\end{abstract}
\begin {IEEEkeywords}
Deep Reinforcement Learning, Pilot Assignment, Pilot Contamination, Massive MIMO. 
\end{IEEEkeywords}
\section{Introduction}
\label{intro}
\IEEEPARstart{T}
{he} ever increasing demand for wireless throughput necessitates  innovative technologies to be developed. Granted the high throughput achieved by large number of antennas, massive multiple-input multiple-output (M-MIMO) is considered as a solution to these requirements. 
By employing time division duplexing (TDD) for transmission and reception,  pilot-based techniques  can be used for channel state information (CSI) acquisition in such systems.
However, in case the users are assigned  correlated pilot sequences, the CSI acquisition faces inevitable interference, referred to as pilot contamination which remains a challenging concern that can be minimized through pilot decontamination techniques. 

In \cite{Zhu2015} a smart pilot assignment (SPA) method is presented in which
pilot assignment is performed such that the least interference is caused for the users with worst channel quality.
To this end, a cost function is formed based on the large-scale fading coefficients of the users, and it is minimized through a sequential algorithm. 
Building on this work, the authors in \cite{Shahabi2019a}  presented low complexity pilot assignment schemes based on the large scale fading coefficients of each user. Although the performance of the algorithms is good, consideration limited to only large scale fading coefficients, which depends on only the distance of users from the BS, may not be enough. Thus, in \cite{Muppirisetty2018} a pilot assignment scheme was presented based on both large-scale fading coefficient and the angle of arrival (AoA) information of the users, resulting in a higher performance compared to the SPA method.
The authors in \cite{Zhu2016} proposed a soft pilot reuse (SPR) method for reducing the pilot contamination effects. Although they showed that the SPR method can increase spectral efficiency in many practical cases, such methods require large value of overhead as a result of employing large number of pilot sequences. 

Recently, learning-based methods have been exploited to address the problem of policy design of pilot assignment.  In \cite{PilotDesign1}, a deep learning-based pilot design algorithm was presented to reduce pilot contamination for a multi-user M-MIMO system. To be specific, through unsupervised learning, a multi-layer fully connected deep  neural network (DNN) was devised to solve the Mean Square Error (MSE) minimization problem online.  
Moreover, The authors in \cite{PilotDesign2} presented a solution for joint optimization of pilot design and channel estimation in order to minimize the MSE of channel estimation in a multi-user MIMO system through introducing a deep learning-based joint pilot design and channel estimation scheme, where  two-layer neural networks (TNNs) is utilized for pilot design, in addition to DNNs for channel estimation. In \cite{PilotDesign3}, employing adaptive moment estimation (ADAM) algorithm, an end-to-end DNN structure was presented to  design the pilot signals as well as the channel estimator, in which a fully-connected layer was devised  through compressing the high-dimensional
channel vector as input to a low-dimensional vector implying the received measurements. Nonetheless, since the proposed DL-base assignment methods follow a blind search trajectory, a huge amount of offline data would be inevitable to cover all possible pilot assignment patterns. 

Motivated by the above, in this paper, we propose an AOA-based pilot assignment scheme for  multi-cell M-MIMO system using deep reinforcement learning (DRL). At first, a cost function representing the pilot contamination is designed based on the location of the users to determine their channel quality. Then, by defining proper sets of states, sets of actions and reward functions based on the channel characteristics and the resultant maximum cost functions,  the agent learns the policy for pilot assignment that can adapt to channel variations while minimizing the cost function. Numerical results show that the proposed method is able to track different channel realizations and select the relevant pilot assignment that results in a close performance to that of the exhaustive search algorithm.


\section{System Model and Problem Formulation}
\label{System Model}
In this paper,  we consider the uplink transmission of a multi-cell M-MIMO system, which consists of $L$ cells each containing a BS with $M$ antennas serving  $K$ single antenna users. The location of the $k$-th user of the $l$-th cell is denoted by $\mb{z}_{kl}$ and the location of the $j$th BS is represented by $\mb{z}_j$.
The channel coefficient between this pair of user and BS is demonstrated by $\mb{g}_{jkl}=\sqrt{\frac{D_{jkl}}{P}}\sum_{p=1}^{P}\mb{a}(\omega_{jkl}^{(p)})\alpha_{jkl}^{(p)}$,
where $D_{jkl}=c\|\mb{z}_{kl}-\mb{z}_j\|_2^{-\eta}$, $\eta$ is the path loss coefficient and $c$ is a constant depending on cell-edge SNR. Moreover, $\mb{a}(\omega_{jkl}^{(p)})\in\mathbb{C}^M$ represents the antenna steering vector corresponding to $\omega_{jkl}^{(p)}\in[0,2\pi)$, in which $p$ is the path index from the set of $P$ possible paths, and $\alpha_{jkl}^{(p)}$ is  the stochastic phase of the $p$-th path. The sonstant $c$ is calculated for the cell radius $R$ as $c[\db]=\gamma_{\snr}+10\eta\log_{10}(R)+10\log_{10}(\sigma^2)$,
where $\gamma_{\snr} (dB)$ denotes the cell edge SNR and  $\sigma^2$ is the noise variance in the receiver. Using the uniform linear arrays (ULA) the antenna steering vectors for each antenna are modeled as $[\mb{a}(\omega_{jkl}^{(p)})]_m=\exp(-j2\pi md\cos(\omega_{jkl})/ \lambda)$,
where $d$ and $\lambda$ represent the distance between the antennas and the signal wavelength respectively. Moreover, $[\mb{f}]_j$ denotes the $j$-th element of the vector $\mb{f}$. Assuming that the number of paths, $P$, tends to infinity, and the AoAs are independent and identically distributed (i.i.d) random variables, using the law of large numbers it can be concluded that the channel vector $\mb{g}_{jkl}$ is with a zero mean Gaussian distribution with the following covariance matrix $ \mb{R}_{jkl}{=}\mathbb{E}\left[\mb{g}_{jkl}\mb{g}_{jkl}^H\right]{=}D_{jkl}\int_{0}^{2\pi}p(\omega_{jkl})\mb{a}(\omega_{jkl})\mb{a}^{H}(\omega_{jkl})\de\omega_{jkl}$,
in which $p(\omega_{jkl})$ represents the probability density function (PDF) of the variable $\omega_{jkl}\in [\omega_{jkl}^{\min},\omega_{jkl}^{\max}]$. In this uniform distribution, the values of $\omega_{jkl}^{\min}$ and $\omega_{jkl}^{\max}$ are calculated by
$\omega_{jkl}^{\min}=\omega_{jkl}^{\mu}-\omega_{jkl}^{\delta}$ and $\omega_{jkl}^{\max}=\omega_{jkl}^{\mu}+\omega_{jkl}^{\delta}$, where $\omega_{jkl}^{\mu}=\arctan\left(\frac{[\mb{z}_{kl}]_2-[\mb{z}_{j}]_2}{[\mb{z}_{kl}]_1-[\mb{z}_j]_1}\right)$ and $\omega_{jkl}^{\delta}=\arcsin\left(\frac{r_{kl}}{[\mb{z}_{kl}]_1-[\mb{z}_j]_1}\right)$ \cite{Muppirisetty2018}.
 Here, $r_{kl}$ represents the scatter radius around the $k$-th user in the $l$-th cell. Also, $[\mb{f}]_{n}$ denotes the $n$th element of the vector $\mb{f}$. 
  Now, we assume that a target user with an AoA within the region $I_{jkj}=[\omega_{jkj}^{\min},\omega_{jkj}^{\max}]$ is located in the system. The goal is to assign pilots to the interfering users in the neighbouring cells in a way that the interference on the channel estimation of the target user would be minimized. The AoA of these users with respect to the target BS is within $I_{jS(l,k^{[j]})l}=[\omega_{jS(l,k^{[j]})l}^{\min},\omega_{jS(l,k^{[j]})l}^{\max}]$. The notation $S(l,k^{[j]})$ stands for a user in the $l$th cell that has the same pilot as the $k$th user in the $j$th cell.  In the Massive MIMO systems it is shown that the channel estimation process for the target user would be without any error if $I_{jkj}$ and $I_{jS(l,k^{[j]})l}$  are non-overlapping. Furthermore, it can be shown that the users with the steering vectors  $\mb{a}(\omega_{jS(l,k^{[j]})l})/\sqrt{M}$, cause only a negligible channel estimation error for  the target user, in case the following value is small \cite{Muppirisetty2018}
  \begin{align}\label{aoa condition}
      &\frac{\mb{a}^H(\omega_{jS(l,k^{[j]})l})\mb{R}_{jkj}\mb{a}(\omega_{jS(l,k^{[j]})l})}{M}=\nonumber\\&\frac{1}{M}\int J^2(\omega_{jkj},\omega_{jS(l,k^{[j]})l})p(\omega_{jkj})\de \omega_{jkj}.
  \end{align}
In (\ref{aoa condition}), the value of  $J(\omega_{jkj},\phi)$ is calculated as $J(\omega_{jkj},\phi)=\sqrt{D_{jkj}}\left|\sum_{m=1}^{M} \exp(2\pi j(m-1)\frac{d}{\lambda}\times\left(\cos(\phi)-cos(\omega_{jkj})\right)\right|$.
Based on this, an alternative method for minimizing (\ref{aoa condition}) is presented in \cite{Muppirisetty2018}, where for pilot assignment a cost function is used which represents the interference of a user in the $l$th cell that has the same pilot sequence as the $k$th user in the $j$th cell. This cost function is represented by \begin{equation}\label{approximated J}
      G_{k^{[j]},S(l,k^{[j]})}=G_{k^{[j]}}^{aprx}(\omega_{jkl}^{\min})+ G_{k^{[j]}}^{aprx}(\omega_{jkl}^{\max}),
  \end{equation}
  in which 
  \begin{align}\label{J-aprx}
      &G_{k^{j}}^{aprx}(\phi)=\sqrt{D_{jkj}}\times\nonumber\\&\left\{
      \begin{array}{ll}
      1, & \cos(\phi)\leq\cos\left(\pi-\omega_{jkj}^{\min}\right), \\
      1-\zeta_1(\phi), & -\cos\left(\omega_{jkj}^{\min}\right)\leq\cos(\phi)\leq\cos(\psi_{jkj}^{\max}), \\ \zeta_2(\phi), & \cos\left(\psi_{jkj}^{\min}\right)\leq\cos(\phi)\leq\cos\left(\omega_{jkj}^{\min}\right), \\ 1, & \cos(\phi)\geq\cos(\omega_{jkj}^{\min}), \\0, & \ew.
      \end{array}\right.
  \end{align}
  In (\ref{J-aprx}), 
  the values for $\zeta_1(\phi)$ and $\zeta_2(\phi)$ are calculated by $\zeta_1(\phi)=\frac{\cos(\phi)+\cos(\omega_{jkj}^{\min})}{\cos\left(\psi_{jkj}^{\max}\right)-cos(\omega_{jkj}^{\min})}$ and $\zeta_2(\phi)=\frac{\cos(\phi)-\cos(\psi_{jkj}^{\min})}{\cos(\omega_{jkj}^{\min})-\cos(\psi_{jkj}^{\min})}$.
  Moreover, $\psi_{jkj}^{\min}$ and $\psi_{jkj}^{\max}$ are determined via calculating the zeros of the functions $J(\omega_{jkj},\phi)$ and $J(\pi-\omega_{jkj},\phi)$
  as \cite{Muppirisetty2018}
  \begin{align}
      \psi_{jkj}^{\min}=\max\left(\min_r\left(\{\phi_{r,\omega_{jkj}^{\min}}\}_r\right),\min_r\left(\{\phi_{r,\pi-\omega_{jkj}^{\min}}\}_r\right)\right),\\
      \psi_{jkj}^{\max}=\min\left(\max_r\left(\{\phi_{r,\omega_{jkj}^{\min}}\}_r\right),\max_r\left(\{\phi_{r,\pi-\omega_{jkj}^{\min}}\}_r\right)\right),
  \end{align}
  where $\{\phi_{r,\omega_{jkj}^{\min}}\}_r$ and $\{\phi_{r,\omega_{jkj}^{\min}}\}_r$ denote the zeros of $J(\omega_{jkj},\phi)$ and $J(\pi-\omega_{jkj},\phi)$ respectively.
  It is noteworthy to mention that as the number of BS antennas tends to infinity, equation (\ref{approximated J}) tends to $\displaystyle\max_{\omega_{jkj}}J(\omega_{jkj},\phi)$. In the following, by using a DRL-based strategy, the cost function in (\ref{approximated J}) is applied to find a near-optimal pilot assignment strategy.
  \section{DRL-Based Pilot Assignment Strategy}

 In this section, at first, the basics of DRL  is explained, then the algorithm is applied to the pilot assignment problem. 
\subsection{DRL Algorithm}

  \textcolor{\revisioncolor}{Any RL system contains an agent  which interacts with the environment in a series of discrete time steps, a set of states and a set of actions. 
  By taking an action in a time step, the agent transitions from one state to the next. Also, each action results in a reward for the agent. 
  The reward function is introduced as $R:\mathbb{S}\times\mathbb{A}\rightarrow\mathbb{R}$, with $\mathbb{S}$, $\mathbb{A}$ and $\mathbb{R}$ representing the discrete set of states, the set of actions and the set of rewards, respectively. Accordingly,  the episode return is defined as 
      $R_t=\sum_{i=1}^T\mu^{i-t}R(a_i,s_i)$.
  Here, $\mu\in[0,1]$ stands for the discount factor of the future rewards, $a_i\in\mathbb{A}$ represents  the action in the $i$th time step and $s_i\in\mathbb{S}$ is the state of the $i$th time step. 
  By taking the action $a_i$ the state of the agent transitions to $s_{i+1}$ and the agent gains the reward $r_{i+1}$.
  Utilizing Q-learning \cite{Sutton2018}, an efficient policy is adopted to estimate the state-action value function (or in other words Q-function) that maximizes the future reward. The value function is represented as the expected return over all episodes, when starting from a state $s$ and performing the action $a$ by following the policy $\pi:\mathbb{S}\rightarrow\mathbb{A}$ as $Q^{\pi}(s,a)=\mathbb{E}[R_t|s_t=s,a_t=a,\pi]$.
  Using Bellman equation, \textcolor{\omissioncolor}{ \cite{Sutton2018},} the optimal value function is given by $Q^{*}(s,a)=\max_{\pi}Q^{\pi}(s,a)$ and the optimal policy follows $\pi^*(s)=\arg\max_{a\in \mathbb{A}} Q^{*}(s,a)$.
  }
  \textcolor{\revisioncolor}{Additionally, in case of a large number of states, deep RL (DRL) is employed as a potential solution to systems suffering from infeasible generalization for unobserved states. 
  To the best of our knowledge, DRL accounts for using neural networks to approximate the Q-function.}
  The outgoing construction referred to as Q-neural network (QNN) leads to the following approximation for the action-value function $q(s,a,\boldsymbol{\kappa})\approx Q^{*}(s,a)$, where $\boldsymbol{\kappa}$ refers to some parameters that define the Q-value.
  In a specific time step $t$, where the state is $s_t$ and the QNN weights are $\boldsymbol{\kappa}$, the DRL agent takes an action with regards to $a_t=\arg_a \max q(s,a,\boldsymbol{\kappa})$ where $q(s,a,\boldsymbol{\kappa})$ is the output of QNN for every possible action $a$. Then, the agent receives the reward $r_{t+1}$ and transitions to the state $s_{t+1}$. Thus, the experience set at the time step $t$ would be $(s_t,a_t,r_{t+1},s_{t+1})$ which is used in training QNN.
  The Q-value $q(s,a,\boldsymbol{\kappa})$ is then updated towards the target value 
  \begin{equation}\label{target x}
    x^{trg}_{r_{t+1},s_{t+1}}=r_{t+1}+\mu \max_{a} q(s_{t+1},a,\boldsymbol{\kappa}).
\end{equation}
The update rule in DRL is to find the value of $\boldsymbol{\kappa}$ in QNN through a training phase, in which the square loss of $q(s,a,\boldsymbol{\kappa})$ is minimized.
  The square loss of $q(s,a,\boldsymbol{\kappa})$ is defined as
  \begin{equation}\label{prediction error}
      w(s_t,a_t,r_{t+1},s_{t+1})=(x^{trg}_{r_{t+1},s_{t+1}}-q(s_t,a_t,\boldsymbol{\kappa}))^2,
  \end{equation}
Furthermore, the values of $\boldsymbol{\kappa}$, are updated by a semi-definite gradient scheme used for minimizing (\ref{prediction error}) as
\begin{equation}\label{Omega update}
  \boldsymbol{\kappa}\leftarrow \boldsymbol{\kappa}+\rho  [x^{trg}_{r_{t+1},s_{t+1}}-q(s_{t},a_t,\boldsymbol{\kappa})] \bigtriangledown q(s_{t},a_t,\boldsymbol{\kappa}),
\end{equation}
in which $\rho$ is the learning rate step size. 
As the QNN weights are updated, the target value changes.
In DQN, the quasi-static target network approach is used in which the target Q-network $q(.)$ in (\ref{target x}) is replaced by $q(s_{t},a_t,\hat{\boldsymbol{\kappa}})$, and the parameter $\hat{\boldsymbol{\kappa}}$ is updated after every $T$ time steps by $\hat{\boldsymbol{\kappa}}=\boldsymbol{\kappa}$.
Moreover,  the experience replay method \cite{ref27TCOM} can be used for stability of the system. In this method, instead of training QNN at the end of each time step with only one experience, multiple jointed experiences can be utilized for batch training.  In other words, a replay memory with a fixed capacity is considered in which the set $v=(s,a,r,s^{'})$ is saved in specific time steps. A mini-batch referred to as $B$, with $M_{\text{batch}}$ random experiences is selected for a training course, and the loss-function is calculated based on them. 
 By using the experience replay method and the quasi-static target network approach, 
the variable $\boldsymbol{\kappa}$ is updated by
\begin{align}
    \boldsymbol{\kappa}\leftarrow \boldsymbol{\kappa}+\frac{\rho}{M_{\text{batch}}}\sum_{b\in B} [x^{trg}_{r,s^{'}}-q(s,a,\boldsymbol{\kappa})] \bigtriangledown q(s,a,\boldsymbol{\kappa}),
\end{align}
where $x^{trg}_{r,s^{'}}=r+\mu \max_{a^{'}} q(s^{'},a^{'},\hat{\boldsymbol{\kappa}})$.
In the following the DRL algorithm is employed for the purpose of pilot assignment. 

  \subsection{Application of DRL Algorithm in Pilot Assignment  }
    In this section, using the cost function in (\ref{approximated J}), a pilot assignment method is proposed. In this case, the  cost function for the $i$th user in the $j$th cell is presented as $G_{k^{[j]}}=\sum_{\substack{l=1\\l\neq j}}^{L}G_{k^{[j]},S(l,k^[j])}$.
Considering this cost function, the optimization problem for pilot assignment becomes $\min_{\bs{\pi}}\quad\left[\max_{k,j} \ \ G_{k^{[j]}}\right]$.
  Note that, this optimization problem  is based on the AoA distribution of the users, because of the definition for $G_{k^{[j]}}$.
  
  To design a DRL platform for solving this problem, first the state set, the action set and the reward set need to be defined.  Hence, the state set in the time step $n$ is defined by the following information:
  \begin{itemize}
      \item The binary pilot assignment pattern for all users.
     \item The pilot index and the cell index of the users with the maximum cost function in each cell, in the  time step $n-1$.
      \item The pilot index of the user that action takes place on, in the  time step $n-1$.
      \item The cell index of the user that action takes place on, in the  time step $n-1$.
      \item  The value of the maximum cost function in each cell in the system, in the time step $n$.
  \end{itemize}
   We assume that the pilot index and the cell index of the user with the maximum cost function are denoted by $k''$ and $l''$, respectively. Also, $k$ and $l$ stand for the pilot index and the cell index of the selected user by the agent, to exchange its pilot with another user in the $l$th cell, with the same pilot as the target user. In this case, the state set can be represented by $\mathbb{S}=\Big\{\Tilde{u}_j(i), G_{[j]}, k, l , k'' , l'' \ \Big| \ j=1,\ \dots,L \ , \ i=1,\dots,K\Big\}$,
where $G_{[j]}$ represents the value of the maximum cost function in the $j$th cell. The notation $\Tilde{u}_j(i)$ stands for the binary index of a user in $j$-th cell which uses the $i$-th pilot. 

In order to introduce the action set, the user with the maximum cost function is considered. For the sake of simplicity we refer to this user as the target user, and its cell as the target cell. Then, in the neighbouring cell, a random user is selected and it’s pilot is switched with the pilot of the user that is assigned with the same pilot sequence as the target user. Note that, in case the selected user already has the same pilot as the target user, no action would be taken. To be more specific, when the DRL structure is moving from the $n$th time step to the $n+1$th time step, the action set for the target user is $\mathbb{A}_{kl}=\{a_{k',l'}|k'=1,\dots,K \ , \ l'=1,\dots,L\}$,
in which $a_{k',l'}$ is defined as
\begin{equation}
    a_{k',l'}=\left\{\begin{array}{ll}
        \text{No action is taken}, &  k'=k\\    \left\{\begin{array}{ll}
     u_{l'}^{(n+1)}(k')=u_{l'}^{(n)}(k)&  \\u_{l'}^{(n+1)}(k)=u_{l'}^{(n)}(k')
             & 
        \end{array}\right.& k'\neq k
    \end{array}\right.
\end{equation}
where $u_{l}^{(n)}(k)$ denotes the user of the $l$-th cell that utilizes the $k$-th pilot sequence at the time step $n$.

The next step is to define the reward set. To do so, two thresholds are considered for the cost function as $g_1$ and $g_2$ ($g_2>g_1$). In this case, by moving from the $n$th to the $n+1$th time step, the instantaneous reward is defined as $r=r^{(1)}+r^{(2)}+r^{(3)}$,
where 
\begin{equation}\label{AoA instantanous reward}
    r^{(1)}=\left\{\begin{array}{ll}
        +1, & G_{[l'']}^{(n+1)}<g_1  \\
         0, & g_1<G_{[l'']}^{(n+1)}<g_2\\
        -1, & G_{[l'']}^{(n+1)}>g_2,
    \end{array}\right.
\end{equation}
\begin{equation}\label{AoA overhead reward}
    r^{(2)}=\left\{\begin{array}{ll}
        -1, & \text{Action is taken}  \\
         0, & \text{Action is not taken},
    \end{array}\right.
\end{equation}
\begin{equation}\label{AoA transition reward}
    r^{(3)}=\left\{\begin{array}{ll}
        +2, & G_{[l'']}^{(n+1)}<g_1, \ G_{[l'']}^{(n)}>g_2  \\
         +1, & G_{[l'']}^{(n+1)}<g_1, \ g_1<G_{[l'']}^{(n)}<g_2\\
         +1, & g_1<G_{[l'']}^{(n+1)}<g_2, \ G_{[l'']}^{(n)}>g_2\\
         -1, & g_1<G_{[l'']}^{(n+1)}<g_2, \ G_{[l'']}^{(n)}<g_1 \\
         -1, & G_{[l'']}^{(n+1)}>g_2, \ g_1<G_{[l'']}^{(n)}<g_2\\
         -2, &G_{[l'']}^{(n+1)}>g_2, \ G_{[l'']}^{(n)}<g_1\\
         0, &\text{Otherwise}
         
    \end{array}\right.
\end{equation}
Note that, $G_{[l'']}^{(n)}$ and $G_{[l'']}^{(n+1)}$  represent $G_{[l'']}$ before and after taking an action in the $n$th time step, respectively. Furthermore, $r^{(1)}$ in (\ref{AoA instantanous reward}) represents the certain reward achieved in the $n+1$th time step, and  $r^{(2)}$ in (\ref{AoA overhead reward}) denotes the cost of overhead on the system caused by taking an action in the $n$th time step. Obviously, taking any action, regardless of the certain reward it might or might not achieve, results in a negative reward due to the overhead it causes for the system. Finally,  $r^{(3)}$ in    (\ref{AoA transition reward}), stands for the relative reward during the transition from the $n$th time step into the $n+1$th time step. Based on this definition for the reward set, it can be claimed that taking any action could result in a negative instantaneous total reward, unless  the certain reward and the relative reward are adequately positive. This definition for the reward set, results in taking more targeted actions.
\section{Numerical Results}
\begin{figure}[t]
\begin{center}
   \includegraphics[scale=.27]{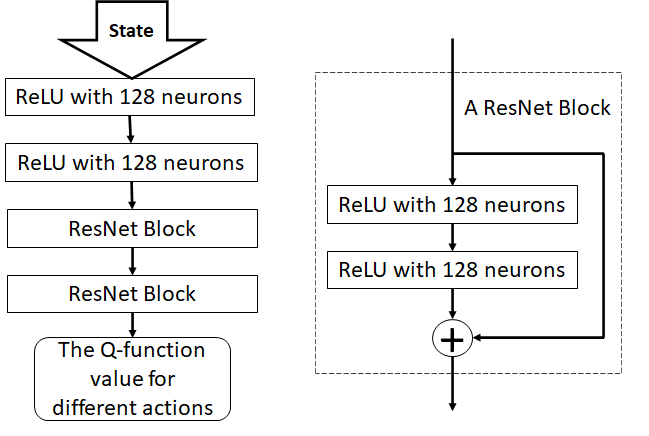}
       \caption{The structure of a QNN realised by ResNet}
    \label{fig:1}
\end{center}
\end{figure}

In this section, the DRL-based algorithm is simulated and analyzed. To evaluate the proposed method, the minimum rate among all users is used as a comparison benchmark. Furthermore, the proposed schemes are compared to the exhaustive search method and the random pilot assignment, which respectively give the upper and the lower bounds of performance for the pilot assignment problem. The performance of the DRL-based scheme is also compared to that of the soft pilot reuse method in \cite{Zhu2016}, which necessitates using more orthogonal pilot sequences than the number of users in each cell, resulting in high system overhead.  It is assumed that the system is composed of $L=7$ cells  each containing a BS with $M=100$ antennas serving $K=4$ single-antenna users.  Also, the  path  loss  coefficient is set to $\eta=2.5$.  The QNN structure utilized for the deep reinforcement learning is a deep residual network (ResNet) \cite{He2016} with six hidden layers. This QNN structure is depicted in Fig. \ref{fig:1}. In this structure, each hidden layer contains 128 neurons. For the sake of simplicity, we refer to the realization of QNN by ResNet as ResNet. Furthermore, each item in this ResNet structure is referred to as a ResNet block. The ReLU functions \cite{Goodfellow} in this structure, are considered as the activation functions for the neurons. The first two hidden layers of this ResNet are completely connected to each  other and they are followed by two  ResNet blocks.    
\begin{figure}[t]
\begin{center}
   \includegraphics[scale=.44]{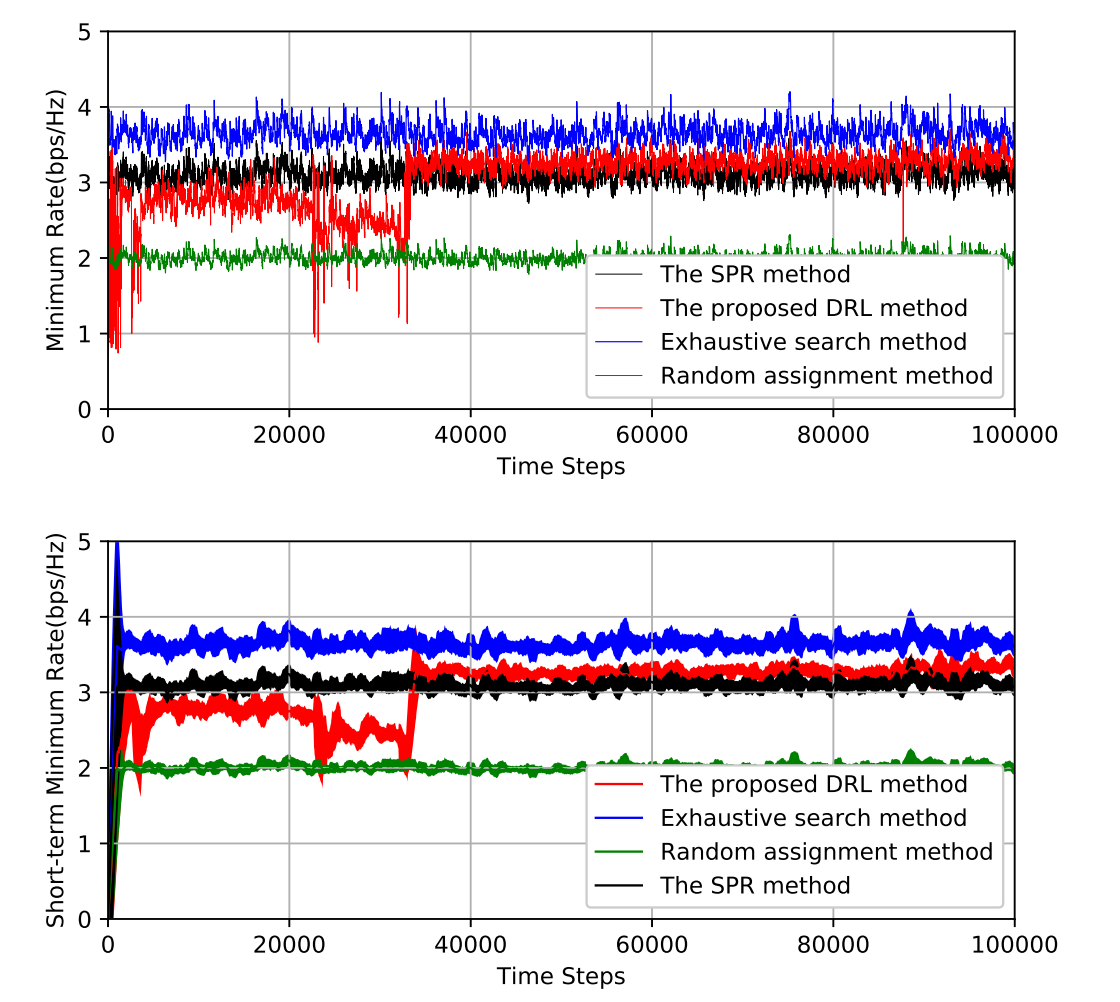}
       \caption{Minimum achievable rate versus versus the AoA time steps; a comparison among the proposed DRL method, the SPR method, the exhaustive search method and the random pilot assignment.}
    \label{fig:2}
\end{center}
\end{figure}
\begin{figure}[t]
\begin{center}
   \includegraphics[scale=.43]{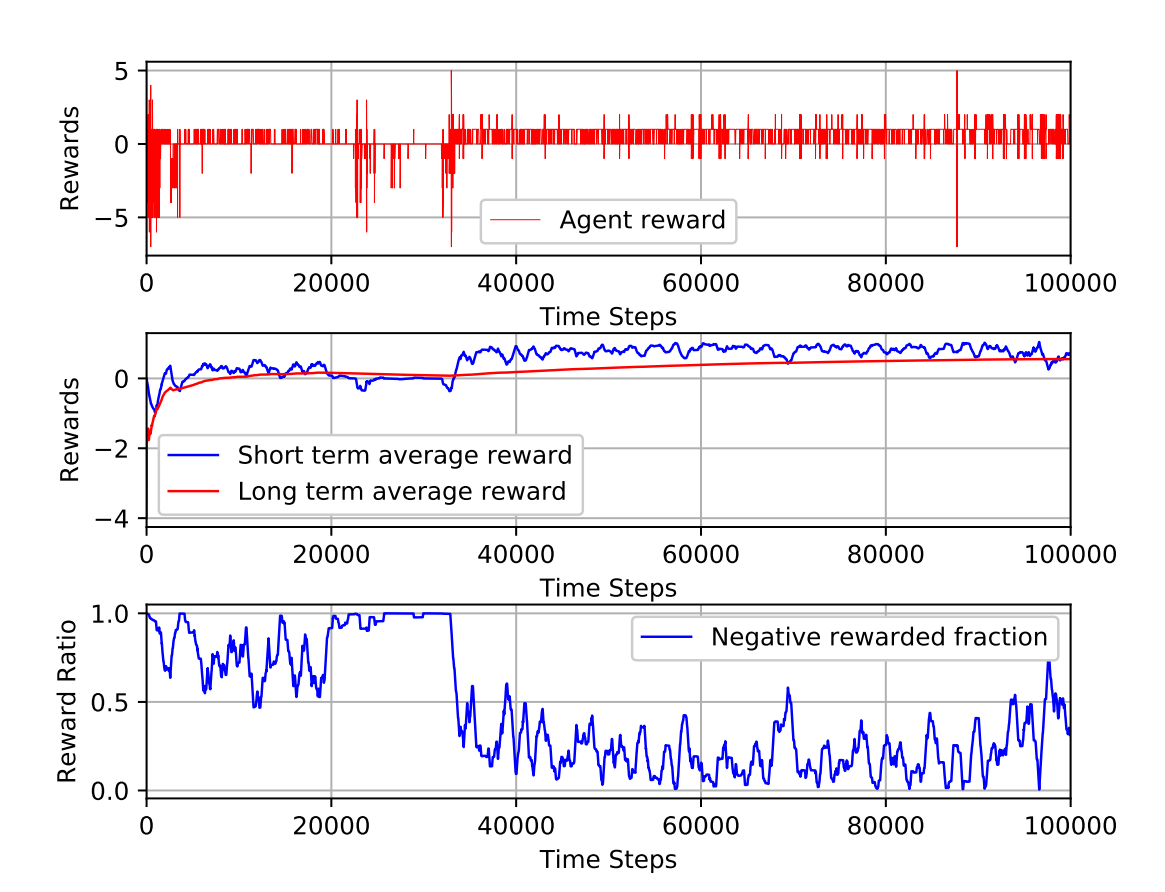}
       \caption{The achievable reward versus the AoA time steps. The First figure demonstrates the agent's rewards throughout time, the second figure shows the short-term and the long-term rewards, and the third figure illustrates the negative reward ratio versus the AoA time steps. }
    \label{fig:3}
\end{center}
\end{figure}
Each ResNet block contains two consecutive hidden layers  plus a shortcut from the input to the output of the ResNet block. Whenever the $\bs{\kappa}$ coefficient in the QNN is updated, a mini-batch with 200 experimental samples are randomly selected from the 500 previous experiences in the experience replay reservoir, in order to calculate the loss function. Note that the experience replay reservoir is updated in a first in first out (FIFO) manner, and whenever the experience memory is full, the older experiences are removed for the new experiences to be restored. Also, in order to update $\bs{\kappa}$ by the mini-batch gradient descent method, the RMSprop \cite{Tieleman2012} algorithm is used. An exponential Decay $\epsilon$-greedy Algorithm is applied in the system, so that the DRL structure would not get stuck in a sub-optimal decision policy before learning adequate experiences. 
The value of $\epsilon$ at first is set to $0.5$, but it is decreased gradually with every time step by the rate of $0.9975$, until it reaches a threshold of $0.0001$. Note that, having a positive value for the $\epsilon$ at all times results in adaptability of the decision policy to the future changes. Also, the discount factor $\mu$ is set to $0.9$.

In Fig. \ref{fig:2}, the performance of the DRL algorithm is depicted in terms of the minimum achievable rate versus the AoA time steps, and it is compared to the exhaustive search method, the random pilot assignment and the SPR scheme. In this figure, to calculate the short-term average minimum rate in each time step, the average of the minimum rates over the last 50 time steps is calculated.
Each time step contains a new channel realization. As the time steps increase, the performance of the proposed method tends to the performance of the exhaustive search algorithm. Ergo, it can be concluded that by the passage of time, the proposed algorithm gains the ability to track the changes in the channels and learn the effective pilot assignment policy. Note that in the higher time steps, the gap between the performances of the proposed DRL method and the exhaustive search algorithm is small. Also, after gaining enough experiences, the proposed scheme out-performs the SPR method while the system overhead of the SPR method is much greater than that of the DRL-based scheme.  To be specific, assuming that in the SPR method,  the ratio of marginal users to central users is 1/3,
in a cluster of 7 cells, the number of required orthogonal pilot sequences of the SPR method
is $250\%$ greater than that of the DRL-based scheme.

Fig. \ref{fig:3} describes the performance of the DRL method in terms of the achievable reward. The first figure demonstrates the agent's rewards throughout time. It can be seen that as the time passes, the agent learn the right policy to achieve mostly positive rewards. In the second figure, to calculate the short-term achievable reward in each time step, the average of the rewards among the last 50 time steps is calculated. Furthermore, to calculate the long-term reward in each time step, the reward in all of the time steps, from the beginning until the current step are considered. As the time step increases, the positive rewards supersede the negative ones by the agent. Moreover, the short-term reward and the long-term reward are both fully positive in the higher time steps.  The third figure, shows the ratio of the negative rewards, and it can be seen that as the agent gains more experiences, the ratio of the negative rewards decreases. Hence, it can be concluded that this algorithm is on the right path and finally it reaches an effective pilot assignment.

\section{Conclusion}
In this paper, the problem of pilot assignment in multi-cell M-MIMO systems is tackled by the DRL scheme. By using both the distance and the AoA information of the users, a cost function is defined representing the pilot contamination effects in the system, and an optimization problem is formed to minimize this cost function. The DRL algorithm is applied to this problem, to find an effective pilot assignment strategy. 
Numerical results show that the performance of the DRL-based scheme is better than some methods in the literature while it maintains a lower system overhead.

\bibliographystyle{IEEEtran}
\bibliography{Mendeley,Ref1,MyRef}
\end{document}